\documentclass[article,showpacs,preprintnumbers,amsmath,amssymb]{revtex4-1}
\usepackage{graphicx}
\usepackage{subcaption}
\usepackage[utf8]{inputenc} 
\usepackage{array}
\begin{document}
\author{M. Haluk Se\c cuk}
\email{haluksecuk@marun.edu.tr}
\affiliation{Physics Department, Faculty of Science and 
	Letters, Marmara University, 34722 Istanbul, Turkey}
\title{Superradiance of a Global Monopole in Reissner-Nordstr\"{o}m(-AdS) Space-time}
\author{\"Ozg\"ur Delice}
\email{ozgur.delice@marmara.edu.tr}
\affiliation{Physics Department, Faculty of Science and 
Letters,  Marmara University, 34722 Istanbul, Turkey}

\date{\today}
\begin{abstract}

In this article, the behavior of a charged and massive scalar field around a global monopole swallowed by a Reissner-Nordstr\"{o}m-Anti-de Sitter (RN-AdS) black hole is investigated by considering the Klein-Gordon equation in this geometry.  The superradiance phenomenon and instability behavior of the black hole against charged scalar perturbations are studied for both an RN-AdS case and also for an RN black hole surrounded by a reflective mirror, i.e., the black hole bomb  case. The effects of the monopole on these cases are discussed  analytically and  also with the help of several graphs in detail. The monopole charge affects the superradiance threshold frequency and also effects the instability time scale for both cases. The existence of global monopole makes these black holes more stable against superradiance instability. 
\end{abstract}
\maketitle

\section{Introduction}

Global monopoles are a special class of topological defects which may be produced in the early universe during the symmetry breaking phase transitions in Grand Unified Theories (GUT) \cite{Kibble,Vilenkin1}. Various types of these defects such as monopoles, cosmic strings, domain walls or textures may be produced depending on the type of the broken symmetry. A possible mechanism for global monopoles to be formed is a spontaneously broken global $O(3)$ symmetry to $U(1)$\cite{referans1}. These monopoles have interesting gravitational properties such as the spacetime around a global monopole has a solid deficit angle and the active gravitational mass of these monopoles vanishes \cite{referans1}, despite very tiny repulsive force due to the core \cite{Harari}. They have also remarkable cosmological implications, such as these monopoles,  unlike magnetic monopoles, are free from monopole  overproduction problem in the early universe. This is because there is a very large attractive force between global monopoles and anti monopoles, which implies that they have very efficient annihilation process and the monopole overproduction problem may not exist \cite{referans1} for these monopoles. Hence, as a result, very few number of monopoles, even none, can be present in the universe per Hubble radius. They have also interesting astrophysical implications such as global monopoles, similar to local cosmic strings, formed at the GUT scale, can serve as  seeds for galaxy and large scale structure formation \cite{Bennet}. However,   observations on cosmic microwave background radiation showed that these topological defects cannot be the main contributor of the density fluctuations giving rise to large scale structures, but they are not completely ruled out and  can contribute as secondary sources, no more than a few percent \cite{Pogosian,Bevis},  to the density fluctuations leading to the structure formation. 
For a review of topological defects and their astrophysical and cosmological implications, we refer to \cite{referans5}. Hence,  understanding the gravitational effects of these defects and their interaction with the surrounding scalar, or electromagnetic fields are still an important topic in theoretical and observational astrophysics and cosmology. In order to increase our knowledge on the properties and behaviour of global monopoles, in this article, we would like to investigate the dynamics of a massive, charged test scalar field in a spacetime where a global monopole is swallowed by a charged, massive black hole, namely, Reissner-Nordstr\"{o}m-AdS(RN-AdS) black hole. We first investigate the Klein-Gordon equation(KGE) of a charged, massive scalar test particle. By using known mathematical methods, we can solve the KGE approximately which enables us to explore the instability conditions of the superradiance phenomenon for these spacetime configurations.

Superradiance \cite{referans10} is a phenomenon present in dissipative systems leading to a radiation amplification process. Examples of this phenomenon for different areas of physics can be found in \cite{BekensteinReview,BritoReview}.   This phenomenon also exists for both rotating  \cite{referans2,zeldovich1,Bardeen,Starobinsky,Starobinsky1} and  charged  black holes \cite{Bekenstein,Denardo,Furuhashi}. In the superradiant scattering, a scalar or electromagnetic wave sent far from the black hole is scattered by the black hole with increased wave strength. This is because the horizon of the black hole, which  acts as a one way membrane, plays the role of a dissipative system, and leads to the superradiant scattering,  if the frequency of the wave is below a certain threshold frequency. 
 The wave is enhanced because it gains rotational or electromagnetic energy from the black hole and therefore it is a wave analogue of the Penrose process. Hence, using this mechanism, one could, in theory, extract energy from the black hole. The area theorem is safe in this process, since,  during this process  not only the  rotational and electromagnetic energy  but  also the angular momentum  and horizon charge of the black hole decreases, in a way that the horizon area does not decrease.  
 One important property of the superradiant scattering is that it may lead to an instability if there exists a mechanism to feed the enhanced scattered waves back into the black hole until the scattered waves exceed the threshold frequency of this black hole. In the classical domain,  the first example of the phenomenon was presented by Zel'dovich \cite{referans2}, whose suggestion was to surround a rotating cylindrical absorbing object by a reflecting mirror, then by examining the case where the scalar waves strike upon it. 
 It is known that Schwarzschild black holes  do not show superradiant scattering and hence they are stable \cite{ReggeWheeler,Vishveswara}. It is possible to obtain a system similar to Zel'dovich's cylinder for a rotating Kerr black hole surrounded by a spherical reflective mirror. This black hole-mirror system is called the "black hole bomb" by Press and Teukolsky  \cite{Press}. A detailed analytical and numerical study investigating superradiance instability of Kerr black holes for the black hole bomb mechanism can be found in \cite{referans3}. These studies showed that using a hypothetical reflective mirror which continuously sends back scattered waves enhanced by superradiant scattering into the black hole, until the mirror breaks down due to the radiation pressure of the scattered waves, one can observe the superradiant instability of a black hole and its behaviour such as the instability time scale. This method is applied to several black hole configurations \cite{zeldovich1,Strafuss,HodHod,Rosa,Witek,Dolan,Hodbomb,Herdeiro,Lee,Dias1,Dias2,DeliceTurkuler,Degollado,Degollado1,Degollado2,Degollado3}  to investigate different aspects of the instability of these systems against superradiance.

 A more important question is that are there any natural mechanisms, for example a potential barrier, which continuously scatters back enhanced waves into the black hole. One such mechanism is  the effect of the mass of the scalar wave,  which  behaves like a potential barrier  \cite{Damour,Zouros,Detweiler,Furuhashi}, prevents scalar waves enhanced by superradiant scattering to escape radial infinity  and reflects back to the black hole for Kerr family of rotating black holes. However, the time scale of this instability for astrophysical black holes are greater than the Hubble time, hence this instability is ineffective for such black holes. However, this instability may be important for primordial black holes. Another candidate is the infinity of AdS black holes since for AdS black holes the infinity behaves like a reflective mirror. Despite this, large AdS black holes were shown to be stable \cite{HawkingReal}. However, four dimensional rotating and/or charged \cite{referans4,Uchikata1, Bosch,Gonzalez,Gubser1,Uchikata}  small  AdS black holes can be unstable against superradiant scattering if the scalar field satisfies the superradiance condition.  This observation is also present in rotating  and/or charged  \cite{AlievDelice,DeliceTurkuler,Wang,Li1,Kodama}  AdS black holes  in higher dimensions. Moreover, gravitational or Maxwell perturbations may also trigger superradiance instability for AdS black holes \cite{Cardoso1,CardosoGrav,WangMaxwell,Basu}. This instability causes the black hole to loose some of its angular momentum and horizon charge.  Finding the end product at the onset of the instability has been a question in recent years,  with the expectation that  a charged rotating AdS  black hole having only a single helical Killing vector with smaller angular momentum and charge surrounded by a possible charged scalar cloud \cite{Dias3,Dias6}   will be the answer. However, a recent study showed that this end product is also unstable \cite{Green}, implying that the answer is not fully reached, yet.

As we have discussed in the previous paragraphs, the superradiance phenomenon is not specific to rotational black holes. For static charged black holes such as RN and RN-AdS black holes, since the charge of the black hole also leads to superradiance \cite{Bekenstein,Denardo,Furuhashi}, it can also be observed  via scattering of  a charged scalar field interacting wih the Coulomb energy of the black hole instead of a rotational one.  Hence, in this case, superradiant scattering reveals itself for the frequencies bounded by the inequality $\omega<e\Phi_h$, where $e$ is the charge of the scalar field and $\Phi_h$ is the electric potential of the horizon of the black hole sourced by its charge $Q$. In this article, we generalize the instability condition in the presence of a global monopole for both RN and RN-AdS spacetimes. Unlike the rotating ones, there are no metastable bound states in the superradiant regime of RN  black holes for charged massive scalar fields, and therefore asymptotically flat charged black holes are stable against superradiance instability \cite{Moncrief1,Moncrief2,Furuhashi,Hod1,Hod2,Menza,Baake}. However, the  superradiant scattering may  lead to an instability if static charged black holes are surrounded by  a reflective mirror \cite{Degollado,Degollado1,Degollado2,Degollado3,Hodbomb,Herdeiro} or lives in an AdS  spacetime \cite{Uchikata,Wang,Li1,Li2}. The endpoint at the onset of the instability as a charged hairy black hole with a charged scalar cloud floating above the horizon  is pursued in linear \cite{Basu,Dias4,Dias5,Gentle} and nonlinear \cite{Bosch} regimes for static charged black holes. 
In order to see the effect of the monopole on the superradiance instability of static charged black holes, we  will analyze the instability in both of  these setups.  For a more complete list of references to  black hole superradiance and the other aspects of superradiance, we refer to the latest review \cite{BritoReview}.

Our aim in this paper is to investigate  the interaction of a global monopole  with a (charged) scalar field via the superradiance phenomena. The questions that we will be seeking answers are the following: What is the effect of the monopole on the superradiance threshold frequency?  How does the monopole affect the superradiance instability? Does the existence of the monopole affect the instability time scale and if it is so,  is it an increase or a decrease? The motivation behind these questions is to understand the characteristic behavior of global monopoles in the presence of a scalar field. In order to answer these questions we consider a  global monopole swallowed by RN(-AdS) black hole. The choice of this geometry is mandatory because of the following two reasons. The first one is the fact that the Klein-Gordon equation is not separable for "rotating global monopole space-time \cite{Filho}" and hence we are left with a non-rotating but charged RN spacetime, which shows superradiance {\bf scattering} property. The second one is, in order to investigate the superradiance {\bf instability}, we need a configuration that has a  "barrier" which  continuously reflects  scattered waves back into  the black hole. This can be achieved by either an AdS boundary which behaves like a reflective mirror, or an artificial setting where the black hole we are considering is surrounded by a reflective mirror. Hence, we will consider both cases separately to compare their results. The third option is to consider a massive scalar field will not be pursued in this paper because the potential barrier due to the mass term does not lead to  instability for  RN black holes. Although   global monopoles that might be produced  in the early universe in the symmetry breaking phase transitions seem to be unrelated to the settings we will be considering in this paper, it may not be true. Once a monopole is produced, a global monopole network shows a scaling property such that a fixed number of order one monopoles survive per  Hubble radius, thus the number of monopoles do not change in  an expanding universe \cite{Bennet,Turok}. Hence  there is a possibility that a global monopole might be swallowed by a black hole in its lifetime. Therefore it is not unreasonable to investigate the properties of a black hole configuration where a global monopole swallowed by an RN(-AdS) black hole and its interaction with a scalar field.       

The article is organized as follows. In section 2 we present the spacetime corresponding to a global monopole  swallowed by  a black hole with mass $M$, and  electrical charge $Q$ in a cosmological background which is called RN-AdS global monopole spacetime and we obtain the Klein-Gordon equation for a scalar field by using an ansatz to separate the equation to its angular and radial parts. The quasinormal modes of RN-dS black hole with a global monopole are presented recently in \cite{Zhang}. Section 3 is devoted to the phenomenon of superradiance. We analytically investigate the stability properties of the spacetime configuration in two different cases under superradiance. In the first case, we have an AdS spacetime which behaves effectively as a reflecting box. In the second case, however, we consider the spacetime in the absence of the cosmological constant and surround the black hole by a reflecting box. For both cases, we discuss the effect of the global monopole on the superradiant instability by using known analytical methods. We also present several graphs to visualize the effect of the monopole on the superradiant threshold frequency and also on the time scale of the instability for both AdS and black hole bomb cases. In section 4 we make a brief conclusion of the results we have found in this paper.\\

\section{The Line Element And The Klein-Gordon\\Equation}\label{sec1}

In this section, we will present the line element for our spacetime configuration and also the Klein-Gordon equation for a charged scalar field. The behavior of the scalar field near the horizon and the radial infinity are derived. Using these results the superradiance condition is also derived and the effect of the monopole charge on the superradiant threshold frequency is discussed.
\subsection{The Line Element}\label{subsec2}

The spacetime line element around a global monopole swallowed by a Schwarzschild black hole is given in \cite{referans1}. This solution is later generalized to a global monopole swallowed by an RN-(A)dS black hole implicitly in \cite{Guendelman} with the line element
\begin{equation}\label{lineelement}
ds^2=-\frac{\Delta_r}{r^2}\,dt^2+\frac{r^2}{\Delta_r}\,dr^2+r^2\left(d\theta^2 + \sin^2\theta\,d\phi^2\right),
\end{equation}
with $\Delta_r$ is defined as,
\begin{equation}\label{Deltar}
\Delta_r=(1- 8\pi\eta^2)r^2- 2Mr-\frac{\Lambda}{3}r^4+{Q^2},
\end{equation}
where $M$ and $ Q$ are the total mass and the total charge of the black hole, $\Lambda$ is the cosmological constant, and $\eta$ is the contribution of the global monopole, which are the physical parameters of this space-time. The fourth degree polynomial \eqref{Deltar} has two real and two imaginary roots for AdS space-time. These two real roots corresponds to the event horizon $r_+$ and Cauchy horizon $r_-$. From now on we will  mainly work with the following parameters
\begin{equation}b^2=(1- 8 \pi \eta^2),\quad \ell=\sqrt{-\frac{3}{\Lambda}}.
\end{equation}
and now the function $\Delta$ can be expressed in terms of the abbreviation term $b$ and the AdS radius $\ell$ as
\begin{equation}\label{Delta1}
\Delta_r=b^2r^2- 2Mr+\frac{r^4}{{\ell}^2}+{Q^2}.
\end{equation}

Hawking temperature of RN-AdS spacetime is presented in  \cite{Berti}, and for RN-AdS-monopole spacetime it becomes
\begin{align}\label{hawkingtemp}
T &= \frac{1}{4\pi}\left(\frac{d}{dr}\frac{\Delta_r}{r^2}\right)\Bigg\rvert_{r=r_+} = \frac{1}{4 \pi  r_+}\left(\frac{b^2}{4 \pi  r_+}+\frac{3 r_+}{4 \pi  \ell^2}-\frac{Q^2}{4 \pi  r_{+}^{3}}\right).
\end{align} 
The positive definiteness of Hawking temperature constraints the charge of the black hole by the  following inequality,
\begin{equation}\label{const:eq}
Q \leq Q_{c} := r_{+} \sqrt{b^2 + \frac{3 r_{+}^{2}}{\ell^2}}.
\end{equation}
which reduces to the corresponding critical charge expression \cite{Berti,Uchikata} in RN-AdS spacetime when $b=1$.
We can see that, the  monopole term definitely effects the Hawking temperature and the critical charge $Q_c$  in a nontrivial way, both via the term $b^2$ and also with its effect on $r_+$.  
In the following sections,  we will choose the black hole parameters such that the constraint equation (\ref{const:eq}) is obeyed.

One important note comes from the inspection of the pure global monopole configuration. The line element for such a configuration can be obtained by neglecting the black hole parameters, namely the mass $M$, the charge $Q$ and $\Lambda$. Therefore the line element becomes \cite{referans1,referans5},
\begin{equation}
ds^2=-(1- 8 \pi \eta^2)\,dt^2+\frac{1}{(1- 8 \pi \eta^2)}\,dr^2+r^2d\theta^2+r^2\sin\theta^2d\phi^2.
\end{equation}
Rescaling the $t$ and $r$ variables by the transformations, 
\begin{equation}
t \rightarrow \frac{t}{\sqrt{1- 8 \pi \eta^2}},\quad  \quad r\rightarrow \sqrt{1- 8 \pi \eta^2}\,r ,
\end{equation}  
we can rewrite the global monopole line element as,
\begin{equation}\label{assympmonmet}
ds^2=-dt^2+dr^2+(1- 8 \pi \eta^2)\,r^2\left(d\theta^2+\sin\theta^2d\phi^2\right).
\end{equation}
The line element \eqref{assympmonmet} not only describes the asymptotic behavior of the global monopole outside the core but also states that the pure global monopole spacetime is not asymptotically flat, which describes a space with a solid deficit angle. Hence the area of a sphere of proper radius $r$ is not $ 4 \pi r^2$, but rather $ (1- 8 \pi \eta^2)\,4 \pi r^2$. Note that the space-time (\ref{assympmonmet}) also describes a "cloud of strings" solution  \cite{Letelier}, namely, a  configuration where an ensemble of radially distributed straight cosmic strings, i.e., a Letellier spacetime, intersecting at a common point, which sometimes also called the "string hedgehog configuration \cite{Guendelman,Delice}".  
Note also that for positive values of $(1- 8 \pi \eta^2)$, i.e. $(1- 8 \pi \eta^2)>0$, the equation \eqref{lineelement} defines a spacetime such that $\Delta_r=0$ at a certain value of r. However for $(1- 8 \pi \eta^2)<0$, 
$r$ becomes a timelike variable and \eqref{assympmonmet} can be interpreted as an anisotropic cosmological solution.

\subsection{The Klein-Gordon Equation}\label{subsec4}

The Klein-Gordon equation for a scalar field $\Phi$ that describes the dynamics of a massive electrically charged scalar particle of mass $\mu$ and charge $e$, in a curved spacetime is described by the equation,
\begin{equation}\label{KG}
\square\Phi=\frac{1}{\sqrt{-g}}D_{\mu}(\sqrt{-g}g^{\mu\nu}D_{\nu})\Phi=\mu^2\Phi,
\end{equation}
where $g$ is the determinant of the metric tensor, with the value $g=-r^4\sin^2\theta$. The gauge differential operator $D$ is defined as,
\begin{equation}
D_\mu=\partial_\mu-ieA_\mu,\quad
A=A_\mu dx^\mu=-\frac{Q}{r}\,dt,\label{electricpotential}
\end{equation}
where $A_\mu$ is the vector potential.

It is straightforward to see that the Klein-Gordon equation is separable. Considering the usual separation ansatz, 
\begin{equation}\label{ansatz}
\Phi=R(r)S(\theta)\,e^{i m \phi}\,e^{-i \omega t}.
\end{equation} 
where $m$ is the azimuthal quantum number and $\omega$ is the angular frequency of the scalar waves, and substituting \eqref{ansatz} into \eqref{KG} yields a separable equation on differential equation \eqref{KG}, which means we can write the total differential equation as distinct  angular and radial equations separately.
The angular part of the total differential equation leads to the associated Legendre differential equation
\begin{equation}
\left[ \frac{1}{\sin\theta}\frac{d}{d\theta}\left(\sin\theta \frac{d}{d\theta} \right)-\frac{m^2}{\sin^2\theta} \right]S(\theta)=\lambda\, S(\theta),
\end{equation}
where $\lambda$ is the separation constant with well known expression
\begin{equation}
\lambda=\nu(\nu+1),
\end{equation}
 whose solutions are given in terms of associated Legendre polynomials as
\begin{equation}
S(\theta)=P_\nu\,^m(\cos\theta),
\end{equation}
  whose values can be found by using the Rodrigues' formula \cite{referans9}.

The radial part of the Klein-Gordon is obtained as,
\begin{equation}\label{radialequation}
\Delta_r \frac{d}{dr}\left[\Delta_r \frac{d}{dr} R(r)\right] +\left\{(\omega r^2- eQr)^2 - \Delta_r \left[\nu(\nu +1) + \mu^2r^2\right]\right\} R(r)=0.
\end{equation}
 Now, let us discuss the asymptotic behaviour of the scalar field near the horizon and the radial infinity for certain values of the parameters of the scalar field and the black hole.

\subsection{The  Asymptotic behaviour of the scalar field} 

The radial  part of the  Klein-Gordon differential equation is given by the equation \eqref{radialequation}, where $\Delta_r$ is given in Eq. (\ref{Delta1}).
%

Consider now a tortoise coordinate transformation defined as,
\begin{equation}\label{tortoisecoord}
\frac{{\rm d}r^*}{{\rm d}r}=\frac{r^2}{\Delta_r}, \quad\quad \bar{R}(r^*)=R(r)\,r,
\end{equation}
then the equation \eqref{radialequation} takes the Schr\"odinger-like form,
\begin{equation}\label{tortoise}
\frac{{\rm d^2}\bar{R}(r^*)}{{\rm d}{r^*}^2}+ V(r^*)\bar{R}(r^*)=0,
\end{equation}
where the effective potential is defined as,

\begin{equation}
V(r^*)=\left(\omega -\frac{e\,Q}{r}\right)^2 - \frac{\Delta_r}{r^2}\left[\frac{\nu(\nu +1)}{r^2} + \mu^2 +\frac{2}{\ell^2}+\frac{2M}{r^3}-\frac{2Q^2}{r^4}\right].
\end{equation}
Now let us investigate the asymptotic behavior of this potential near the horizon and at the radial infinity.
\subsection{Scalar field near the horizon}
Near the horizon,   $r \rightarrow r_+$, which corresponds to the largest root of $\Delta_r$, the function $\Delta_r$ behaves as $\Delta_r \rightarrow 0$, the effective potential becomes,
\begin{equation}
V(r^*) \rightarrow \left(\omega - e\Phi_{h}\right),\quad\, {\rm as} \quad\quad r \rightarrow r_+, \quad\quad \Delta_r \rightarrow 0,
\end{equation}
Here $\Phi_{h}$ is the electric potential at near the event horizon given by,
\begin{equation}
\Phi_{h}=\frac{Q}{r_+}.
\end{equation}
Hence near the horizon we 
have,
\begin{equation}\label{nhsoln}
\Phi \sim e^{-i\omega t \pm i\left(\omega-e\,\Phi_h\right)r^*},
\end{equation}
where $r^*$ is the tortoise coordinate given by the equation \eqref{tortoisecoord}. Since our investigation is  in the classical domain, we have to choose the negative sign in (\ref{nhsoln}) which implies that there are only ingoing waves at the horizon, and also that one must restrict the group velocity of the wave packet to a negative one. Classically speaking, no information can come out from a static black hole.

\subsection{Scalar field at the infinity}
At the radial infinity there are different asymptotic behaviour for the scalar field depending on the cosmological constant and mass of the scalar field. Hence we will discuss  below these different cases, separately.
\subsubsection{Nonvanishing Cosmological constant case}
For nonvanishing cosmological constant, i.e.,  $\ell\neq \infty$, we have
\begin{equation}
V(r^*)\rightarrow \infty,\quad {\rm as} \quad\quad r \rightarrow \infty\quad ( {\rm for}\quad \ell\neq \infty)
\end{equation}
 which implies that the boundary condition for the scalar field in this case  is the following,
\begin{equation}
R \rightarrow 0 \quad\quad {\rm when} \quad\quad r^* \rightarrow \infty,
\end{equation}
due to the fact that AdS space behaves effectively as a reflecting mirror. 
\
\subsubsection{Vanishing Cosmological Constant case}
For the vanishing cosmological constant, however, the behaviour of the scalar field is very different, since
\begin{equation}
V(r^*)\rightarrow \omega^2-b^2\mu^2,\quad {\rm as} \quad\quad r \rightarrow \infty\quad ({\rm for }\quad\ell= \infty).
\end{equation} 
Hence, for vanishing cosmological constant, and if the scalar field is massive ($\mu\neq 0$), then  bound states that are decaying at infinity are possible for the scalar field if $\omega^2<b^2\mu^2$ with 
\begin{equation}
\bar{R}(r^*) \rightarrow  e^{-\sqrt{b^2\mu^2-\omega^2}\,r^*} \quad\quad {\rm when} \quad\quad r \rightarrow \infty.
\end{equation}
Hence, similar to RN or Kerr black holes, the mass of the scalar field can act as a potential barrier if it satisfies $\omega^2<b^2\mu^2$. We see that the effect of the monopole term is to reduce the height of the potential barrier by a factor of $b^2=1-8\pi \eta^2<1.$ However, it was shown in \cite{Moncrief1,Moncrief2,Furuhashi,Hod1,Hod2,Menza,Baake}  that, unlike Kerr black holes, in the superradiant regime there are no metastable bound states for  RN solution and RN black holes are stable against charged scalar perturbations. Hence we will not pursue the investigation of stability due to the mass of the scalar field in this paper. An open problem will be to investigate the stability of a  global monopole swallowed by a charged and rotating black hole, a solution which awaits its discovery, against charged and massive scalar perturbations. However, as far as we know, this solution is not known yet.
 
 For the case where the mass of the scalar field vanishes or $\omega^2\ge b^2\mu^2$,  then there is no bound state solutions and the field behaves as
\begin{equation}
R(r^*)\rightarrow e^{\pm i\omega_0 r^*}
\end{equation} 
 where $\omega_0=\sqrt{\omega^2-b^2\mu^2},$ with $\omega^2\ge b^2\mu^2$. In this  case  the superradiance scattering cannot lead to instability unless one uses some artificial mechanisms such as surrounding the black hole with a reflective mirror  as done in the black hole bomb mechanism. 
 \subsection{Superradiance  Condition \label{superradiancecondition}}
 Here we derive superradiance condition for vanishing cosmological constant case. Let us consider a scattering experiment of a monochromatic scalar wave with frequency $\omega$ with a wave function of the form $\Phi=\bar{R}e^{-i\omega t+i m \phi}$. When a scalar wave is sent from radial infinity with unit amplitude, and when we consider the black hole horizon as a one way membrane with no flux outside the horizon from the black hole,  then the asymptotic form of the solution of the equation (\ref{tortoise}) can be written as 
\begin{equation}
\bar{R} \sim \left\{  \begin{array}{ll}
\mathcal{T} e^{-i(\omega-e\Phi_h) r^*} & \mbox{as $r\rightarrow r_h$} \\
\mathcal{R} e^{i \omega_0 r^*}+ e^{-i \omega_0 r^*} & \mbox{as $r\rightarrow \infty$}
\end{array} \right.
\end{equation}
Here $\mathcal{R}$ and $\mathcal{T}$ are the amplitudes of the reflected and transmitted waves, respectively. Note that the complex conjugate of $\bar R$, in which we will denote as $ \bar{R}^\dag$, should be also a solution of the equation (\ref{tortoise}) since the potential $V(r^*)$ is real and the solutions are invariant under $t\rightarrow -t$, $\omega \rightarrow -\omega$. Then $\bar{R}$ and $\bar{R}^\dag$ should be linearly independent and their Wronskian  $W=\bar{R} \partial r^*\bar{R}^\dag-\bar{R}^\dag \partial r^*\bar{R}$ should be independent of $r^*$. Calculating Wronskians near horizon and at the radial infinity and equating them one  obtains
\begin{equation}
|\mathcal{R}^2|=1-\frac{\omega-e\Phi_h}{\omega_0}|\mathcal{T}|^2.
\end{equation}
Hence when the superradiant condition 
\begin{equation}
\omega < e\Phi_h
\end{equation} 
is satisfied, then the amplitude of the scattered wave becomes greater than it is sent.   This phenomenon is called as the superradiant scattering. Note that for the AdS case, the condition for superradiance is also the same. This can be derived by the fact that the phase velocity of the waves flowing into the horizon changes sign relative to the  group velocity of these waves. Now let us discuss the role of the monopole charge on the superradiant threshold frequency
\begin{equation}
\omega_p=e \Phi_h=e \frac{Q}{r_+}.
\end{equation}
The monopole  term, $b^2=1-8\pi\eta^2<1$, affects the superradiance threshold frequency since it changes the  location of the outer horizon $r_+$ which is
given by
\begin{equation}
r_+=\frac{M+\sqrt{M^2-Q^2b^2}}{b^2}
\end{equation}
When the monopole is present, the location of the outer horizon increases relative to the case where the monopole is not present ($b=1$). Hence the electric potential of the horizon decreases in the presence of the monopole term. Therefore,  here we conclude that the presence of the monopole charge reduces the superradiant threshold frequency of the wave. A wave with frequency $\omega$ which may trigger the superradiant scattering when the monopole term is absent, may not trigger the superradiant scattering when the monopole term is present. 
  
\section{Superradiance Instability}\label{sec5}

Having obtained the Klein-Gordon equation for  a charged and massive scalar field around a Reissner-Nordstr\"{o}m-(A)dS global monopole black hole in the previous section, and also determined the role of the monopole term on the superradiant threshold frequency,  we now  investigate the phenomenon of superradiance against perturbations of  charged and massive  scalar field to understand the role of the monopole term on the  superradiant instability of the  black holes that we consider. The aim of this section is to find an instability condition for our space-time configuration for small mass and charge via solving the radial wave equation \eqref{radialequation} in the low-frequency domain, i.e. $(r-r_+)<<\frac{1}{\omega}$, by exploiting the asymptotic matching technique. We separate our investigation into two cases where the first case corresponds to the anti-de Sitter spacetime where $\Lambda<0$, that can be called natural superradiant instability since the infinity of the AdS spacetime behaves like a reflective mirror. In the second case, we will be interested in the superradiant instability in the absence of cosmological constant, i.e. $\Lambda=0$, using the method called as the black hole bomb mechanism, where the black hole is surrounded by a hypothetical reflective mirror \cite{Press}.
\subsection{Case 1\,:\,Superradiant Instability of Global Monopole Configuration in  RN-AdS Space-Times\,($\Lambda<0$)} \label{sec5.1} 
In this section we consider the superradiance instability of a charged scalar field of a global monopole swallowed by an RN-AdS black hole. 
As we have said before,  here we exploit the asymptotic matching technique, which divides the spacetime outside the  event horizon into  near and far regions  \cite{Starobinsky}.
\subparagraph{A - Near Region Solution}\

For small AdS black holes we have $r_+<<\ell$, in the near region we assume $(r-r_+)<<\frac{1}{\omega}, \Lambda \sim 0, r \sim r_+$ and $\Delta_r \sim \Delta$, where
\begin{gather}
\Delta=b^2r^2-2Mr+Q^2=(r-r_+)(r-r_-),\quad
r_{\pm}=\frac{M\pm\sqrt{M^2-Q^2b^2}}{b^2}.
\end{gather}
We further assume that $\mu^2r_+^2<<1$ in the near region, since we are in the low frequency regime and the Compton wavelength of the perturbations must be large compared to the radius of the horizon.
Now we will make a change of variable through the following definition,
\begin{align}\label{zvariable}
z=\frac{r-r_+}{r-r_-},\quad\quad  0 \leq z \leq	1,	
\end{align}
where $z=0$ corresponds now the event horizon $r=r_+$. Using \eqref{zvariable} we have the following results,
\begin{align}
\Delta \partial_r=(r_+ - r_-)z\,\partial_z,\quad\Delta=z(r-r_-)^2,\quad(1-z)=\frac{r_+-r_-}{r-r_-}.
\end{align}
 The radial equation \eqref{radialequation} takes the form,
\begin{equation}\label{zvarradial}
(1-z)z\,\partial^2_z\,R+(1-z)\partial_z\,R+\left\lbrace\overline{\omega}^{2}\,\frac{1-z}{z}-\frac{\nu(\nu+1)}{1-z}\right\rbrace R=0,
\end{equation}
where we have defined the so-called superradiant factor as,
\begin{equation}\label{superratiantfactor}
\overline{\omega}=\frac{r_+^2}{r_+-r_-}\left(\omega-e\Phi_h\right).
\end{equation}
Now we can define the following  transformation
\begin{equation}\label{zhomotrans}
R=z^{i\,\overline{\omega}}\left(1-z\right)^{\nu+1}F.
\end{equation}
Substituting \eqref{zhomotrans} to \eqref{zvarradial} we obtain,
\begin{align}\label{hyperradial}
(1-z)z\,\partial^2_z\,F+\left\lbrace(1+2\,i\,\overline{\omega})-\left[2(\nu+1)+2\,i\,\overline{\omega}+1\right]z\right\rbrace \partial_z \, F
+ \left[(\nu+1)^2+(\nu+1)2\,i\,\overline{\omega}\right]F =0,
\end{align}
which is a hypergeometric differential equation with a general solution in the neighbourhood of $z=0$  as \ $F= a z^{1-\gamma}\,F(1+\alpha-\gamma,\beta+1-\gamma;2-\gamma;z)
+b\,F(\alpha,\beta;\gamma;z)
$ \cite{referans9}, where
\begin{equation}
\alpha=\nu+1+2\,i\,\overline{\omega},\quad\quad\beta=\nu+1,\quad\quad\gamma=2\,i\,\overline{\omega}+1.
\end{equation}
Therefore, we can read of the solution of \eqref{zvarradial} as,
\begin{align}\label{nearso}
R=A\,z^{-i\,\overline{\omega}} (1-z)^{\nu+1}\,F(1+\alpha-\gamma,1+\beta-\gamma;2-\gamma;z)+B\,z^{i\,\overline{\omega}} (1-z)^{\nu+1}F(\alpha,\beta;\gamma;z).
\end{align}
Since we are in the classical limit, there will not be outgoing waves, therefore we have to set the coefficient $B=0$.

Now we analyse for the large values of $r$, i.e. $z \rightarrow 1$, the behaviour of the ingoing wave solution in the near region. To accomplish that, we will use hypergeometric transformation law  $z \rightarrow 1-z $, which is given by \cite{referans9},
\begin{align}
F(1+\alpha-\gamma,1+\beta-\gamma;2-\gamma;z)&=(1-z)^{\gamma-\alpha-\beta}\frac{\Gamma(2-\gamma)\Gamma(\alpha+\beta-\gamma)}{\Gamma(\alpha-\gamma+1)\Gamma(\beta-\gamma+1)}F(1-\alpha,1-\beta-\gamma;\gamma-\alpha-\beta;1-z)\nonumber \\
&+\frac{\Gamma(2-\gamma)\Gamma(\alpha+\beta-\gamma)}{\Gamma(1-\alpha)\Gamma(1-\beta)}F(1+\alpha-\gamma,1+\beta-\gamma;\alpha+\beta+1-\gamma;1-z).
\end{align}

Since in the limit $z \rightarrow 1 \Rightarrow 1-z \rightarrow 0$, we can use the property of the hypergeometric function $F(\alpha,\beta;\gamma;0)=1$, to write the large $r$ limit of the near region solution of the form
\begin{equation}
R \sim A\,\Gamma(1-2\,i\,\overline{\omega})\left[\frac{(r_+-r_-)^{-\nu}\,\Gamma(2\nu+1)}{\Gamma(\nu+1)\Gamma(\nu-2\,i\,\overline{\omega}+1)}\, r^\nu +\frac{(r_+-r_-)^{(\nu+1)}\,\Gamma(-2\nu-1)}{\Gamma(-\nu)\Gamma(-2\,i\,\overline{\omega}-\nu)}\, r^{-(\nu+1)}
\right].
\label{nearregionsolution}
\end{equation}

\subparagraph{B - Far Region Solution}\

In the far region we assume $r-r_+>>M$, such that the physical parameters of the black hole, namely the mass
 and the charge can be neglected, i.e. $M\sim0,Q\sim0$. Hence the polynomial \eqref{Deltar} now becomes,
\begin{equation}
\Delta\sim r^2\left(b^2+ \frac{r^2}{\ell^2}\right),
\end{equation}
thus the radial part of Klein-Gordon equation \eqref{radialequation} can be written as,
\begin{equation}\label{farradial}
\left(b^2+ \frac{r^2}{\ell^2}\right)\partial^2_r R+2r\left(\frac{b^2}{r^2}+ \frac{2}{\ell^2}\right)\partial_r R+\left[\frac{\omega^2}{b^2+\frac{r^2}{\ell^2}}-\frac{\nu(\nu+1)}{r^2}-\mu^2\right]R =0.
\end{equation}
Note that the equation \eqref{farradial} is the radial wave equation for AdS space-time with a global monopole. Moreover, we also observe that the monopole term  $b^2$ in equation \eqref{farradial} does not vanish, which is adequate due to the fact that the monopole spacetime is not asymptotically flat. Hence we
must keep the monopole term $b^2$ in the far region approximation.
Let us start our calculation with a coordinate transformation defined as $y=b^2 +\frac{r^2}{\ell^2}$, then we further transform that with $y=b^2x$.
With these transformations equation \eqref{farradial} takes the following form,
\begin{equation}\label{transfarradial}
(1-x^2)x\,\partial_x^2 R +\left(1- \frac{5x}{2}\right)\partial_r- \left\lbrace\frac{\tilde{\omega}^{2}{\ell^2}}{4x}+\frac{\lambda(\lambda+1)}{4(1-x)}-\frac{\mu^2\ell^2}{4}\right\rbrace R=0.
\end{equation}
Here we have set $\tilde{\omega}^2=\omega^2/b^2,\,\lambda(\lambda+1)=\nu(\nu+1)/b^2.$

Now let us use the following definitions,
\begin{equation}
\beta_1=\frac{\tilde{\omega}^2 {\ell^2}}{2},\quad\quad\beta_2=\frac{\lambda}{2}.
\end{equation}
and the  following ansatz,
\begin{equation}\label{faransatz}
R=x^{\beta_1}(1-x)^{\beta_2}F.
\end{equation}
Substitution of \eqref{faransatz} to \eqref{transfarradial} yields,
\begin{equation}\label{hyperrad}
(1-x)x\,\partial_r^2 F+\left[\,\gamma^\prime-x(\alpha^\prime+\beta^\prime+1)\,\right]\partial_x F-\alpha^\prime\beta^\prime F=0  ,
\end{equation}
where we have defined,
\begin{align}
\alpha^\prime&= \beta_1+\beta_2+\frac{3}{4}+\frac{1}{4}\sqrt{9+4\mu^2\ell^2}=\frac{\tilde{\omega} \ell}{2}+\frac{\lambda}{2}+\frac{3}{4}+\frac{1}{4}\sqrt{9+4\mu^2\ell^2},
\\
\beta^\prime&= \beta_1+\beta_2+\frac{3}{4}-\frac{1}{4}\sqrt{9+4\mu^2\ell^2}=\frac{\tilde{\omega} \ell}{2}+\frac{\lambda}{2}+\frac{3}{4}-\frac{1}{4}\sqrt{9+4\mu^2\ell^2},
\\
\gamma^\prime&=\tilde{\omega}\ell+1,
\end{align}
such that,
\begin{align}
\alpha^\prime\beta^\prime=
\left(\tilde{\omega}+\lambda+\frac{3}{2}\right),\quad1+\alpha^\prime+\beta^\prime=2(\beta_1+\beta_2)+\frac{5}{2}.
\end{align}
The equation \eqref{hyperrad} is in the form of hypergeometric differential equation and this equation admits a solution in the neighbourhood of $x=\infty$ as \cite{referans9},\,$F(\alpha^\prime,\beta^\prime;\gamma^\prime;x)= C\,x^{-\alpha\prime}F(\alpha^\prime,\alpha^\prime-\gamma^\prime+1;\alpha^\prime-\beta^\prime+1;\frac{1}{x})
+D\,x^{-\beta^\prime}F(\beta^\prime,\beta^\prime-\gamma^\prime+1;\beta^\prime-\alpha^\prime+1;\frac{1}{x}),
$ hence we can write a solution of \eqref{transfarradial} via \eqref{faransatz} as,
\begin{align}
R=(1-x)^{\beta_2}x^{\beta_1} \left\lbrace C\,x^{-\alpha\prime}F(\alpha^\prime,\alpha^\prime-\gamma^\prime+1;\alpha^\prime-\beta^\prime+1;\frac{1}{x})
+Dx^{-\beta^\prime}F(\beta^\prime,\beta^\prime-\gamma^\prime+1;\beta^\prime-\alpha^\prime+1;\frac{1}{x})\right\rbrace.
\label{farsol}
\end{align}
Taking the limit  $x\rightarrow\infty$ and using $F(\alpha,\beta;\gamma;0)=1$, we see that the solution behaves as,
\begin{equation}
R\sim(-1)^{\beta_2}\left[C\,x^{\frac{-1}{4}\left(3+\sqrt{9+\mu^2\ell^2}\right)}+D\right].
\end{equation}
However, at infinity, AdS spacetime behaves like a wall such that the scalar field $\Phi$ vanishes. This implies the restriction that the coefficient $D$ must vanish.

To explore the equation \eqref{farsol} corresponding to the small values of $r$, i.e. $x\rightarrow1$, we
use the $\frac{1}{x} \rightarrow 1-x$ transformation law of the hypergeometric functions \cite{referans9}, which is given by, 
\begin{align}
F(\alpha,\alpha-\gamma+1;1-\beta+\alpha;\frac{1}{x})=\, x^{\alpha-\gamma+1}&(x-1)^{\gamma-\alpha-\beta}\frac{\Gamma(\alpha-\beta+1)\Gamma(\alpha+\beta-\gamma)}{\Gamma(\alpha-\gamma+1)\Gamma(\alpha)}F(1-\beta,1-\alpha;\gamma-\alpha-\beta;1-x)\nonumber\\
&+\frac{\Gamma(2-\gamma)\Gamma(\alpha+\beta-\gamma)}{\Gamma(1-\alpha)\Gamma(1-\beta)}\,x^\alpha\,F(\alpha,\beta;\alpha+\beta+1-\gamma;1-x).
\end{align}
Note that, when $x\rightarrow1$ we have $x-1 \rightarrow \frac{r^2}{\ell^2 b^2}$. Therefore the far region solution for small values of r is given by,
\begin{align}
R \sim C\,\Gamma(\alpha^\prime-\beta^\prime+1)\left[(-1)^{\frac{\lambda}{2}}\frac{\Gamma(\gamma^\prime-\alpha^\prime-\beta^\prime)}{\Gamma(1-\beta^\prime)\Gamma(\gamma^\prime- \beta^\prime)}\, \left(\frac{r}{\ell b}\right)^\lambda+(-1)^{\frac{-3\lambda}{2}}\frac{\Gamma(\alpha^\prime+\beta^\prime-\gamma^\prime)}{\Gamma(\alpha^\prime)\Gamma(\alpha^\prime-\gamma^\prime+1)}\, \left(\frac{r}{\ell b}\right)^{-\lambda-1}
\right].
\label{farregion1}
\end{align}
We observe that the global monopole term, $b^2=1- 8 \pi \eta^2$ effects the far region solution \eqref{farregion1} as a constant multiple of $r$, therefore we can safely apply the boundaries of the pure AdS space-time to analyse \eqref{farregion1}. When $r \rightarrow 0$ the equation \eqref{farregion1} diverges due to $r^{-\lambda-1} \rightarrow \infty$. To obtain regular solutions we impose the condition as follows,
\begin{equation}\label{regularitycondition}
\Gamma(\alpha^\prime-\gamma^\prime+1)= \infty \quad\quad{if}\quad\quad \Gamma(-m)=\infty,\quad m\in \mathbb{Z}_{+}.
\end{equation}
Thus, the regularity condition \eqref{regularitycondition} enables us to interpret $m$, which takes the values from the nonnegative integer numbers $\mathbb{Z}_{+}$, as a principal quantum number. Hence we obtain the discrete spectrum
\begin{equation}\label{spectrum}
\omega=\frac{2\,b}{\ell}(m+\sigma).
\end{equation}  
For the sake of abbreviation we have defined $\sigma=\lambda/2 +3/4+ \sqrt{9+4\mu^2 \ell^2}/4$. Notice that the result \eqref{spectrum} reduces to given in \cite{referans4} when the mass of the field and monopole term vanishes, i.e. $\mu=0,b^2=1$.

Now, it is natural to assume that the condition \eqref{spectrum} can be interpreted as the generator of the frequency spectrum of the normal modes at large distances, due to the fact that at infinity the structure of the RN-AdS black hole is similar to pure AdS background. In addition, one can still observe the effect of the global monopole in \eqref{spectrum}. Having said that, however, we should approach the current predicament more cautiously, since the inner boundaries of the pure AdS or  RN-AdS black hole spacetimes are very different. For a pure AdS space-time, we have $r=0$ as the inner boundary, on the other hand for the black hole case we have $r=r_+$.  Hence if one wishes to observe the effect of the black hole on the frequency spectrum, one must take into account the possibility of tunneling of the wave through the potential located at $r=r_+$, into the black hole and scattered back. Furthermore, the amplitude of the scattered wave may decrease or grows exponentially and may also cause the superradiant instability. To sum up, the quasinormal mode frequencies for the black hole case can be modified with additional complex frequencies as follows,
\begin{equation}\label{theQN}
\omega_{QM}=\frac{2\,b}{\ell}\left(m+\sigma\right)+i\delta_{AdS}.
\end{equation}
where $\delta_{AdS}$ is possibly a small parameter signalling the effects of the charged  black hole having the global monopole.

Exploiting the assumption \eqref{theQN}, and using the gamma function relations for small $\delta_{AdS}$, we have
\begin{equation}\label{forfar1}
\frac{1}{\Gamma(\alpha^\prime)\Gamma(\alpha^\prime-\gamma^\prime+1)}\simeq i\,(-1)^{m+1}\frac{m!}{(m+\lambda+\epsilon-1)!} \frac{\ell}{2b}\delta_{AdS},\quad\delta_{AdS}<<1,
\end{equation}
where $\epsilon=(3+\sqrt{9+4\mu^2\ell^2})/2$, and
\begin{equation}\label{forfar2}
\Gamma(1-\beta^\prime)\Gamma(\gamma^\prime-\beta^\prime)=\Gamma(-\lambda-m-1/2)\Gamma(m+1+\sqrt{9+4\mu^2\ell^2}/2).
\end{equation}
Using \eqref{forfar1} and \eqref{forfar2} in the  far region solution given by the equation \eqref{farregion1} we obtain,
\begin{align}
R \sim C\,\Gamma(\alpha^\prime-\beta^\prime+1)&\left[\frac{(-1)^{\lambda/2}\Gamma(-\lambda-1/2)(\ell\,b)^{-\lambda}}{\Gamma(-\lambda-m-1/2)\Gamma(m+1+\sqrt{9+4\mu^2\ell^2}/2)}\, r^\lambda\right. \nonumber
\\
&\left.+i\,\delta_{AdS}\,(-1)^{-3\lambda/2+m+1}\frac{\Gamma(\lambda+1/2)}{2}\frac{\ell^{\lambda+2}\,b^\lambda\, m!}{(m+\lambda+\epsilon-1)!}\, r^{-\lambda-1}\right].
\end{align}
Now if we want to be successful at the asymptotic matching procedure of the near region and the far region solutions we need a restriction on $\lambda$. The relation between $\lambda$ and $\nu$ is given by,
\begin{equation}\label{apprxlamda}
\lambda(\lambda+1)=\frac{\nu(\nu+1)}{b^2},
\end{equation}
where $b^2=1- 8 \pi \eta^2$. Taylor expansion of $1/b^2$ yields,
\begin{equation}
\frac{1}{b^2}=1+\mathcal{O} (\eta^2),
\end{equation}
neglecting the $\mathcal{O} (\eta^2)$ term we have $\lambda=\nu.$
Therefore we can write the far region solution as,
\begin{equation}\label{farregionsolution}
R\sim C\,\Gamma(\alpha^\prime-\beta^\prime+1)\left[A^\prime\,r^\nu+ i\,\delta_{AdS}\,B^\prime\,r^{-\nu-1}\right],
\end{equation}
where we set the coefficients of $r^\nu$ and $r^{-\nu-1}$ to $A^\prime$ and $B^\prime$ respectively.

Matching the near region solution \eqref{nearregionsolution} with far region solution \eqref{farregionsolution} yields then,
\begin{align}
\delta_{AdS}\sim(-2\,i)(-1)^{m+1}\frac{\ell^{-2(\nu+1)}b^{-2\nu}}{\Gamma(\nu+1/2)}\,\frac{(r_+-r_-)^{2\nu+1}}{\Gamma(m+1+\sqrt{9+4\mu^2\ell^2}/2)}\,\frac{\Gamma(-2\nu-1)}{\Gamma(-\nu)} \nonumber 
\\ \times
\frac{\Gamma(\nu+1)}{\Gamma(2\nu+1)}\,\frac{\Gamma(\nu+1-2\,i\,\overline{\omega})}{\Gamma(-\nu-2\,i\,\overline{\omega})}\,\frac{\Gamma(-\nu-1/2)}{\Gamma(-\nu-1/2-m)}\,\frac{\left[m+\nu+\left(1+\sqrt{9+4\mu^2\ell^2}\right)/2\right]!}{m!}.
\end{align}
Using the gamma function property given by,
\begin{equation}
\Gamma(1+x)=x\,\Gamma(x)
\end{equation}
we obtain,
\begin{equation}\label{deltavalue}
\delta_{AdS}\sim\ -\xi(\omega- e\,\Phi_h)\,\frac{r_+^2(r_+-r_-)^{2\nu}}{\ell^{2(\nu+1)}\,b^{2\nu}\sqrt{\pi}},
\end{equation}
where we have defined $\xi$ as,
\begin{align}
\xi\equiv\,\frac{(\nu!)^2\left[m+\nu+\left(1+\sqrt{9+4\mu^2\ell^2}\right)/2\right]!}{m!\,(2\nu+1)!\,(2\nu)!}\,
\frac{2^{\nu+2-m}(2\nu+1+m)!!}{(2\nu+1)!!(2\nu-1)!!} \left[\prod_{s=1}^{\nu} (s^2+4\,\overline{\omega}^2)\right][\Gamma(m+1+\sqrt{9+4\mu^2\ell^2})]^{-1},
\end{align}
with $\overline{\omega}=\left[r_+^2/(r_+-r_-)\right]\left(\omega-e\,\Phi_h\right)$ and $\omega=(2b/\ell)(m+\sigma)$. Now we have,
\begin{equation}\label{deltaprop}
\delta_{AdS}\propto-\left(\Re[{\omega_{QM}}]-\,e\,\Phi_h\right).
\end{equation}
Hence the superradiance condition is
\begin{align}
\Re[{\omega_{QM}}]<\,e\,\Phi_h=e \frac{Q}{r_+},\quad\quad{for}\quad\quad \delta_{AdS}>0.\label{superradiantcondition}
\end{align}
The scalar field has dependence of $\omega_{QM}$ as,
\begin{equation}\label{finalresult}
\Phi\propto e^{i\,\omega_{QM}\,t}=\,exp\left[-i\,\Re{(\omega_{QM})}t+\delta_{AdS} t\,\right].
\end{equation}
The equation \eqref{finalresult}, together with the condition \eqref{superradiantcondition} implies that the amplitude of the scalar field grows exponentially and causes instabilities. However one should bear in mind the effect of the global monopole term $b^2$. The relevant  physical choice of the global monopole term is $b^2=1-8\pi\eta^2>0$. Furthermore, if we took $\eta^2$ as a positive number, i.e. $\eta^2>0$, then $0<b^2<1$. To observe the net effect of the global monopole let us write an explicit version of \eqref{superradiantcondition} as,
\begin{equation}\label{opendelta}
\frac{2\,b}{\ell}(m+\sigma)<e\,\frac{Q}{r_+},\quad r_{+}=\frac{M+\sqrt{M^2-Q^2b^2}}{b^2}.
\end{equation}

\begin{figure}[htp]
\centering
    \includegraphics[width=0.65\linewidth]{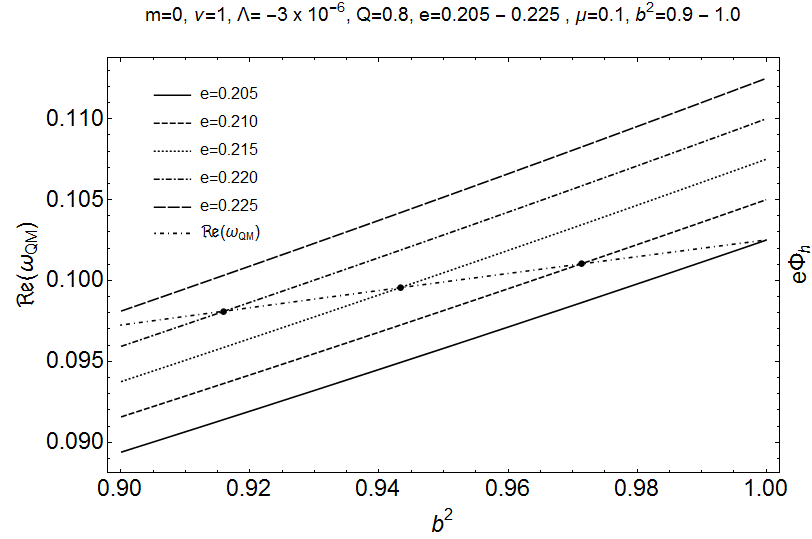} 
  
  \caption{Graph of $e\Phi_h$  and $\Re(\omega_{QM})$ as a function of $
  b^2$ for $m=0$, $\nu=1$, $\Lambda=-3\times10^{-6}$, $Q=0.8$, $\mu=0.1$, $e=0.205-0.220$ with different charge values of scalar field. As $b^2$ becomes smaller, which corresponds to larger values of $\eta$, we see that the condition $\delta_{AdS}>0$ starts not to hold. Instability condition only holds above the intersection points for these choosen parameter
values. Note that, the black hole charge value $Q=0.8$ is in accordance with \eqref{const:eq}.} 
    \label{fig1:SupCond} 
\end{figure}

As we have discussed in section (\ref{superradiancecondition}), we know that the monopole term causes an augmentation on the outer horizon  which decreases the value of the electric potential and lower the threshold frequency. Moreover, inspection of equation \eqref{deltavalue} regarding the effect of global monopole shows a growth in $\delta_{AdS}$ therefore we observe a decrease in $\tau_{AdS}$, since in superradiant instability, the time scale is given by $\tau_{AdS}=1/ \delta_{AdS}=1/\Im[{\omega_{QM}}]\propto\ell^{2(\nu+1)}b^{2\nu}$. 

\begin{figure}[htp] 
  \begin{subfigure}[b]{0.5\linewidth}
    \includegraphics[width=0.99\linewidth]{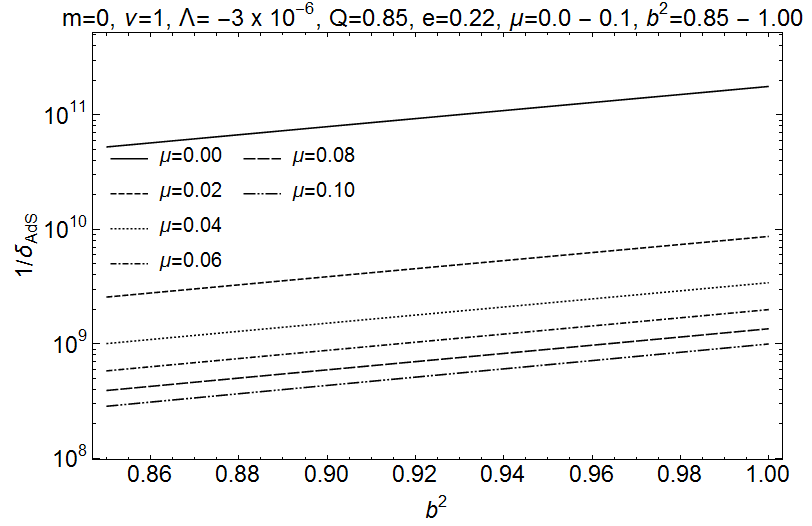} 
    \caption{Logarithmic graph of $\tau_{AdS}$ as a function of $b^2$ for $\mu=0-0.1$, where we have fixed the other parameters as $m=0$, $\nu=1$, $\Lambda=-3\times10^{-6}$, $Q=0.8$, $e=0.22$. i)As the mass of the scalar field decreases, the time scale increases.
     ii) For small values of $b^2$ the time scale decreases. iii) As the mass of the scalar field decreases the slope of the time scale increases.   } 
    \label{fig2:mass:a} 
    \vspace{4ex}
  \end{subfigure}
  \begin{subfigure}[b]{0.5\linewidth}
    \includegraphics[width=0.99\linewidth]{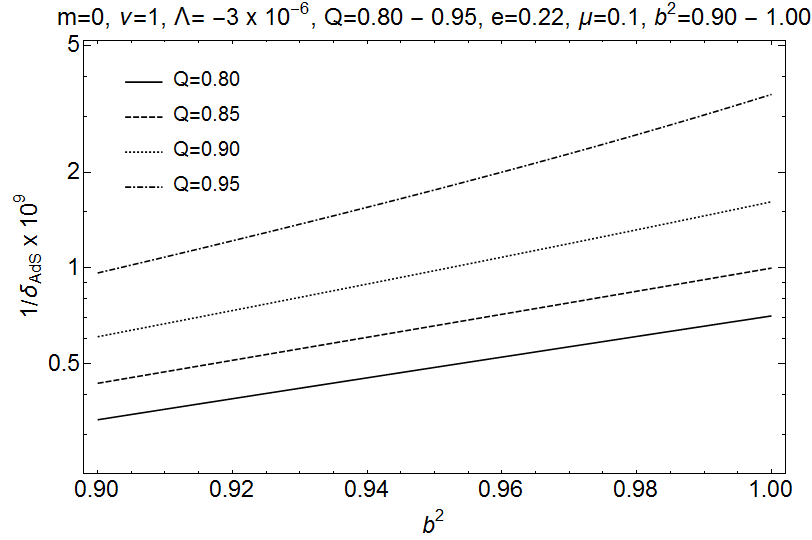} 
    \caption{Logarithmic graph of $\tau_{AdS}$ as a function of $b^2$ for $Q=0.8-0.95$ where we have fixed the other parameters as $m=0$, $\nu=1$, $\Lambda=-3\times10^{-6}$, $\mu=0.1$, $e=0.22$.  i)As the charge of the black hole increases, the time scale increases. ii) As $b^2$ deacreases the time scale decreases. iii) The slope of the time scale increases for increasing $b^2$.} 
    \label{fig2:charge:b} 
    \vspace{4ex}
  \end{subfigure} 
  \begin{subfigure}[b]{0.5\linewidth}
    \includegraphics[width=0.99\linewidth]{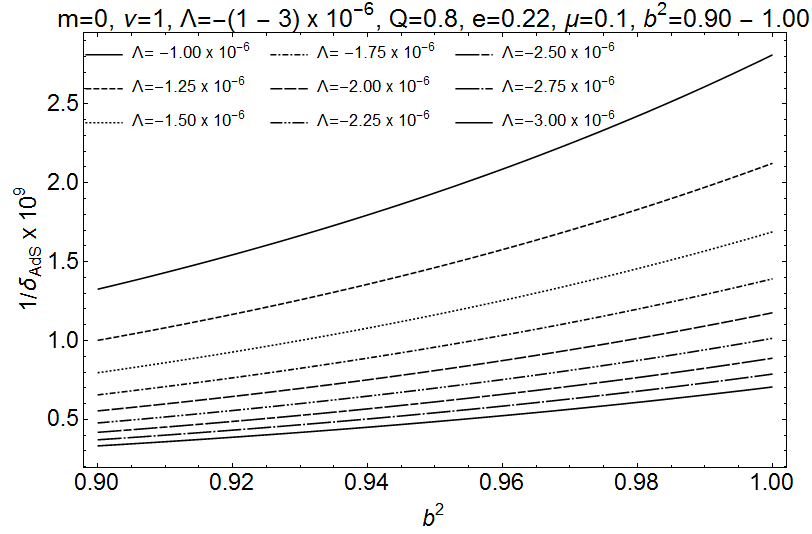} 
    \caption{Graph of $\tau_{AdS}$ as a function of $b^2$ for $\Lambda=-(1-3)\times10^{-6}$ where we have fixed the other parameters as $m=0$, $\nu=1$, $Q=0.8$, $\mu=0.1$, $e=0.22$. i)As the absolute values of the negative cosmological constant increases, the time scale increases. ii) As $b^2$ decreases the time scale decreases. iii) The slope of the time scale increases for increasing $b^2$.} 
    \label{fig2:cosmconst:c} 
  \end{subfigure}
  \begin{subfigure}[b]{0.5\linewidth}
    \includegraphics[width=1\linewidth]{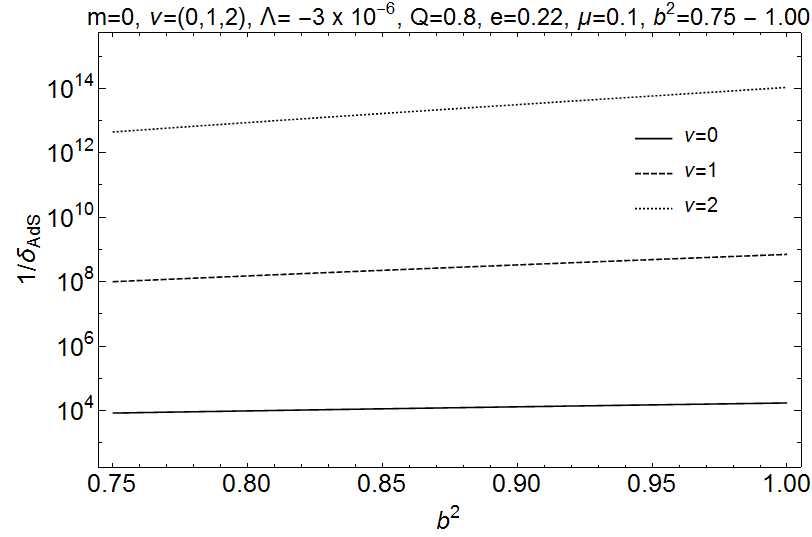} 
    \caption{Logarithmic graph of $\tau_{AdS}$ as a function of $b^2$ for the first three modes $\nu=0, 1, 2$. The straight line corresponding to the \\fundamental mode, i.e., $\nu=0$, shows a little  change compared with the higher order modes. For example when $\nu=2$, the change is in the order of $10^{2}$ whereas for $\nu=0$ the order is practically the same for all values of $b^2$.} 
    \label{fig2:nodes:d} 
  \end{subfigure} 
  \caption{Figures (\ref{fig2:mass:a}), (\ref{fig2:charge:b}), (\ref{fig2:cosmconst:c}), (\ref{fig2:nodes:d}) corresponds to the different values of particle mass, black hole charge, cosmological constant and modes respectively, shows the change of the time scale with respect to $b^2$. All four figures behave in the same fashion, as the effect of the gravitational monopole grows we observe that the time scale decreases. The values of the parameters in the graph are chosen such that the superradiance condition is satisfied. Note that, black hole charge values  in these graphs are lower than the corresponding values of $Q_{c}$ given in Eq. (\ref{const:eq}).}
  \label{fig2:AdSdiff:parameters} 
\end{figure}

In order to better understand the effect of the monopole, now, we will present several graphs. All of the graphs are plotted for unit black hole mass $M=1$.  From figure (\ref{fig1:SupCond}), it is seen that the monopole term $b^2$ plays an important role in the superradiant threshold frequency. As we have said before, when the monopole term is present,  the threshold frequency decreases. As a result, when the black hole contains a global monopole,  the chances of having  a superradiant scattering decreases with increasing monopole term $\eta$. We also plot the graphs  of changing of  the time scale with monopole term $b^2$ for different parameters, namely, the mass of the scalar field $\mu$, the black hole charge $Q$, the cosmological constant $\Lambda$ and different mode values $\nu$ in figures (\ref{fig2:mass:a}), (\ref{fig2:charge:b}), (\ref{fig2:cosmconst:c}), (\ref{fig2:nodes:d}), respectively. We observe that for all of these four parameters, when the gravitational monopole term $b^2$ decreases, the value of the time scale also decreases. In figure (\ref{fig2:mass:a}), when the mass of the scalar field is absent, the time scale differs drastically relative to a massive scalar field. Furthermore, the maximum change of the time scale with respect to gravitational monopole  is observed in this case, as well. Figures (\ref{fig2:charge:b}) and (\ref{fig2:cosmconst:c}) shows the change of the time scale for different values of black hole charge and cosmological constant. As the charge of the black hole and the AdS radius $\ell$ increase, or cosmological constant decrease, the effect of the monopole becomes more obvious. For a given $\mu, Q$ or $\ell$, when monopole term $b^2$ decreases or equivalently $\eta^2$ increases,  the instability time scale  decreases, making the instability less effective for a monopole having more strength. The last figure (\ref{fig2:nodes:d}) is devoted to the  dependence of the time scale to the modes $\nu$ of the scalar wave. The effect of the global monopole on the different  modes of the  perturbation term $\delta_{AdS}$ is given in equation (\ref{deltavalue}). Hence its effect on the instability time scale $\tau_{AdS}$ comes from two different contributions. The first one is the explicit $b^{-2\nu}$ term in $\delta_{AdS}$,  which comes from the coupling between AdS radius $\ell$ and monopole term $b^2$. However, this term is not the only source of $b^2$ dependence  for modes in $\delta_{AdS}$ and as we have said before, the locations of the horizons also have $b^2$ dependence and therefore the term  $(r_+-r_-)^{\nu}$ also depends on mode number $\nu$.  It turns out that the combined effect of these two contributions determines the $b^2$ dependence of  modes to the instability time scale.  The investigation of the figure (\ref{fig2:nodes:d})  for first three modes shows that the main contribution on the time scale change for decreasing $b^2$ comes from the horizon terms since the global monopole effects and increases the values of them.

Hence, we conclude that in RN-AdS black holes having a global monopole, the onset of superradiant instability decreases with the monopole term $b^2$. Nevertheless, if the instability occurs it will grow slower in comparison with the case when the monopole term is absent. In summary, we can conclude that the existence of global monopole makes the RN-AdS black holes more stable against superradiance instability.

\subsection{Case 2\,:\,Black Hole Bomb and Superradiant \\ Instability of Global Monopole Configuration in RN Space-Times\,(\,$\Lambda=0$)}

In this section, we discuss the instability condition in the absence of the cosmological constant $\Lambda$.
As before, we will use the asymptotic matching technique to obtain the instability condition in addition to the so-called mirror condition which will become clear in the process of calculation. Inspection of the near region solution yields the same equation with the AdS case, since we have neglected the effect of the cosmological constant in the case 1 for the near region solution. Hence the near region solution of both cases is the same and we will use the same solution given in equation (\ref{nearso}) and  also we employ its  far region limit given in equation (\ref{nearregionsolution}).
Hence all that remains is to find the far region solution. In the far region, as before, we assume $M\sim0, Q\sim0$, where $M$ and $Q$ is the mass and the charge of the black hole. The polynomial $\Delta_r$ now becomes $\Delta\sim b^2 r^2$. Thus the  radial part of the Klein-Gordon equation \eqref{radialequation} is given by,
\begin{equation}\label{farregdiffc2}
\partial_{r}^2 R+\frac{2}{r}\partial_r R+\left\lbrace \varpi^2-\frac{\lambda(\lambda+1)}{r^2}\right\rbrace R=0,
\end{equation}
where $\varpi^2=\omega^2/b^4-\mu^2/b^2$ and ${\nu(\nu+1)}/ b^2=  \lambda(\lambda+1) $. Equation \eqref{farregdiffc2} admits a general solution in terms of the Bessel functions \cite{referans9} given by,
\begin{equation}\label{Besselsol}
R=r^{-1/2}\left[\alpha\,J_{\nu+1/2}(\varpi\,r)+\beta\,J_{-\nu-1/2}(\varpi\,r)\right],
\end{equation}
and for small values of $r$ it reduces to \cite{referans9},
\begin{equation}\label{smallrBesselsol}
R\sim\frac{\alpha\,(\varpi/2)^{\lambda+1/2}}{\Gamma(\lambda+3/2)}\,r^\lambda+\frac{\beta\,(\varpi/2)^{-\lambda-1/2}}{\Gamma(-\lambda+1/2)}\,r^{-\lambda-1}.
\end{equation}
 Applying the similar mechanical steps that we have performed for the matching procedure in the previous case, we obtain  the corresponding condition for the equations \eqref{smallrBesselsol} and \eqref{nearregionsolution} given as,
\begin{equation}\label{result2}
\frac{\beta}{\alpha}=2\,i\,\overline{\omega}\,\frac{(-1)^\nu}{(2\nu+1)}\,\left(\frac{\nu!}{(2\nu-1)!!}
\right)^2\,\frac{(r_+-r_-)^{2\nu+1}}{(2\nu)!(2\nu+1)!}\left[\prod_{k=1}^{\nu} (k^2+4\,\overline{\omega}^2)\right](\varpi)^{2\nu+1},
\end{equation}
where $b^2=1-8\pi\eta^2$ and $\overline{\omega}$ is the superradiant factor given by the equation \eqref{superratiantfactor}. Notice that, we have used the approximation \eqref{apprxlamda} in order the matching to work. In addition, we have found the coefficient of the near region solution $A$, as the following 
\begin{equation}
A=\alpha\,\frac{(r_+-r_-)^{\nu}}{\Gamma(\nu+3/2)}\,\frac{\Gamma(\nu+1)}{\Gamma(2\nu+1)}\,\frac{\Gamma(\nu-2\,i\,\overline{\omega}+1)}{\Gamma(1-2\,i\,\overline{\omega})}\,\left(\varpi/2\right)^{\nu+1/2}\, ,
\end{equation}
to obtain the \eqref{result2}.

The main difference between the cases is the fact that in case 1 we have an AdS space-time which behaves effectively as a reflecting box. In case  2, however, we put a reflecting mirror by hand at the far region located at a radius $r=r_0$, and as a result, the scalar field must vanish at the surface of the mirror. Hence, we have an additional condition between the amplitudes $\alpha$ and $\beta$ due to the fact that equation \eqref{Besselsol} vanishes for $r=r_0$. Therefore we have, 
\begin{equation}
\frac{\beta}{\alpha}=-\frac{J_{\nu+1/2}(\varpi\,r_0)}{J_{-\nu-1/2}(\varpi\,r_0)},
\end{equation}
and for small values of particle mass $\mu^2<<1$ it yields,
\begin{equation}
\frac{\beta}{\alpha}=-\frac{J_{\nu+1/2}(r_0\,\omega/b^2)}{J_{-\nu-1/2}(r_0\,\omega/b^2)},
\end{equation}
Recalling equation \eqref{result2} we obtain,
\begin{align}\label{mirrorsolution1}
\frac{J_{\nu+1/2}(r_0\,\omega/b^2)}{J_{-\nu-1/2}(r_0\,\omega/b^2)}=\,i\,(-1)^{\nu+1}\,\overline{\omega}\,\frac{2}{2\nu+1}\left[\frac{\nu!}{(2\nu-1)!!}\right]^2 \frac{(r_+-r_-)^{2\nu+1}}{(2\nu)!(2\nu+1)!}
\left[\prod_{k=1}^{\nu} (k^2+4\,\overline{\omega}^2)\right]\left(\frac{\omega}{b^2}\right)^{2\nu+1}.
\end{align}
As a solution to equation  \eqref{mirrorsolution1}, we use the approximations $\omega\ll 1$ and 
$\Re(\omega)\gg \Im(\omega)$, which is adequate for our problem. With these approximations, the $R.H.S$ of the equation \eqref{mirrorsolution1} can be safely set to zero, the result is therefore,
\begin{equation}\label{RHSofmirorsolution1}
J_{\nu+1/2}(r_0\,\omega/b^2)=0,
\end{equation}
which has real solutions \cite{referans9}. We can label the solutions of \eqref{RHSofmirorsolution1}
as,
\begin{equation}\label{labeltoRHS}
j_{\nu+1/2,\,s}=\frac{\omega\,r_0}{b^2},
\end{equation}
where $s$ is a non-negative integer, i.e. $s\in \mathbb{Z}_{+}$. As a complete solution to \eqref{RHSofmirorsolution1} assume that,
\begin{equation}
\omega\sim\ \frac{b^2}{r_0}\left[j_{\nu+1/2, \,s}+i\,\tilde{\delta}\right],\quad\quad \tilde{\delta}<<1.
\end{equation}
Hence, under these assumptions, we have
\begin{align}\label{mirrorsolution2}
\frac{J_{\nu+1/2}(j_{\nu+1/2, \,s}+i\,\tilde{\delta})}{J_{-\nu-1/2}(j_{\nu+1/2, \,s}+i\,\tilde{\delta})}=\,i\,(-1)^{\nu+1}\,\overline{\omega}\,\frac{2}{2\nu+1}\left[\frac{\nu!}{(2\nu-1)!!}\right]^2\frac{(r_+-r_-)^{2\nu+1}}{(2\nu)!(2\nu+1)!}
\left[\prod_{k=1}^{\nu} (k^2+4\,\overline{\omega}^2)\right](\omega/b^2)^{2\nu+1}.
\end{align}
The Taylor expansion of the $L.H.S$ of the equation \eqref{mirrorsolution2} for small values of $\tilde{\delta}$ gives,
\begin{equation}\label{apprxmir}
\frac{J_{\nu+1/2}(j_{\nu+1/2, \,s}+i\,\tilde{\delta})}{J_{-\nu-1/2}(j_{\nu+1/2, \,s}+i\,\tilde{\delta})}\sim i\,\tilde{\delta}\frac{J^{\prime}_{\nu+1/2}(j_{\nu+1/2, \,s})}{J_{-\nu-1/2}(j_{\nu+1/2, \,s})}.
\end{equation}
The values of the expression in the $R.H.S$ of the equation \eqref{apprxmir} can be found in \cite{referans9}. Due to the presence of the mirror located at $r=r_0$, the frequencies of the scalar field therefore are,
\begin{equation}\label{dicretespecofc2}
\omega_{BQN}\simeq\,\frac{b^2}{r_0}j_{\nu+1/2,\,s}\,+\,i\,\delta,
\end{equation}
where $\delta= b^2\tilde \delta/r_0.$ Substitution of \eqref{apprxmir} to \eqref{mirrorsolution2} with \eqref{dicretespecofc2} yields the following result for $\delta$,
\begin{equation}
\delta=-\vartheta\,(-1)^\nu \,\frac{J_{-\nu-1/2}(j_{\nu+1/2, \,s})}{J^{\prime}_{\nu+1/2}(j_{\nu+1/2, \,s})}\,
\frac{\left(j_{\nu+1/2,\,s}\,b^2/r_0\right)-e\,\Phi_h}{r_0^{2\nu+2}}\, b^{2},
\end{equation}
where,
\begin{equation}
\vartheta\equiv\frac{2}{2\nu+1}\left[\frac{\nu!}{(2\nu-1)!!}\right]^2 \left(\frac{r_+^2(r_+-r_-)^{2\nu}}{(2\nu)!(2\nu+1)!}\right)
\left[\prod_{k=1}^{\nu} (k^2+4\,\overline{\omega}^2)\right]\left(j_{\nu+1/2,\,s}\right)^{2\nu+1}.
\end{equation}
Therefore we have,
\begin{equation}\label{mirrorresult}
\delta\propto-\left(\Re[{\omega_{BQN}}]-\,e\,\Phi_h\right)
\end{equation}
and the superradiance condition becomes
\begin{equation}\label{opendeltaBHB}
\Re[\omega_{BQN}]=\frac{b^2}{r_0}j_{\nu+1/2,\,s}<\,e\,\frac{Q}{r_+},
\end{equation}
where the condition $\Re[{\omega_{BQN}}]<\,e\,\Phi_h$ corresponds to  positive values of $\delta$, i.e. $\delta>0$ and as a result the scalar field $\Phi$ grows exponentially and causes an instability. Note that for large values of $r_0$, we obtain small values of $\delta$, hence the assumption $\Re(\omega)>>\Im(\omega)$  remains valid.
\begin{figure}[htp] 
	\begin{subfigure}[b]{0.5\linewidth}
		\centering
		\includegraphics[width=0.95\linewidth]{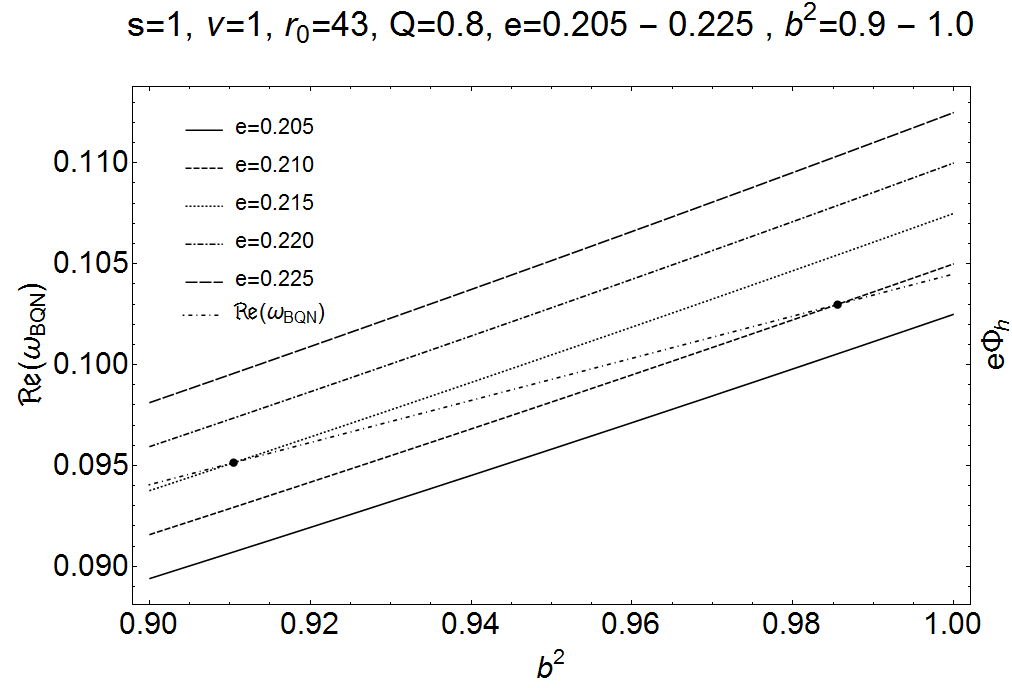} 
		\caption{Graph of $e\Phi_h$ and $\Re(\omega_{BQN})$ as a function of $b^2$ for $s=1$, $\nu=1$, $r_0=43$, $Q=0.8$, with different particle charge values $e=0.205-0.220$. As $b^2$ becomes smaller we observe that the condition $\delta>0$ starts not to hold as in the RN-AdS-Monopole case. Instability condition is only satisfied above the intersection points for these choosen parameter
			values. } 
		\label{fig3:BHBcond:a} 
		\vspace{4ex}
	\end{subfigure}
	\begin{subfigure}[b]{0.5\linewidth}
		\centering
		\includegraphics[width=0.95\linewidth]{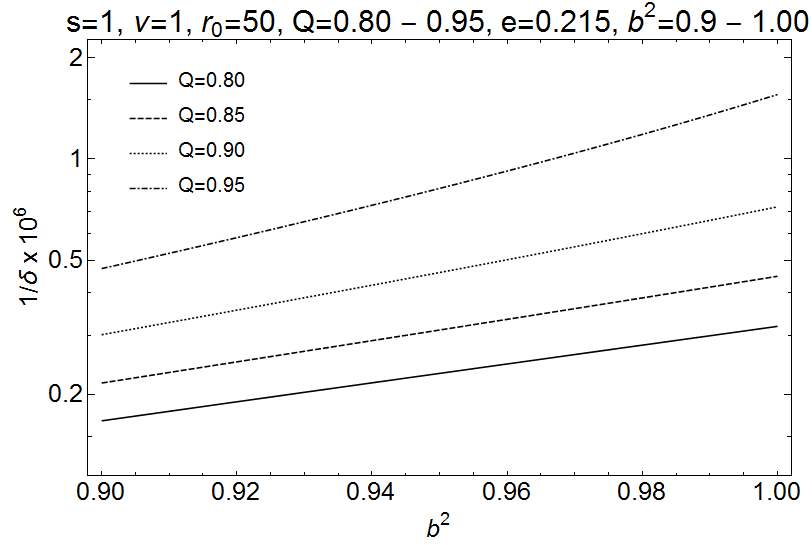} 
		\caption{Logarithmic graph of $\tau$ as a function of $b^2$ for $Q=0.80-0.95$ where we fixed the other parameters as $s=1$, $\nu=1$, $r_0=50$, $e=0.215$. i)As the charge of the black hole increases, the time scale increases. ii) As $b^2$ deacreases the time scale decreases. iii) The slope of the time scale increases for increasing $Q$ and $b^2$.} 
		\label{fig3:BHBcharge:b} 
		\vspace{4ex}
	\end{subfigure} 
	\begin{subfigure}[b]{0.5\linewidth}
		\centering
		\includegraphics[width=0.95\linewidth]{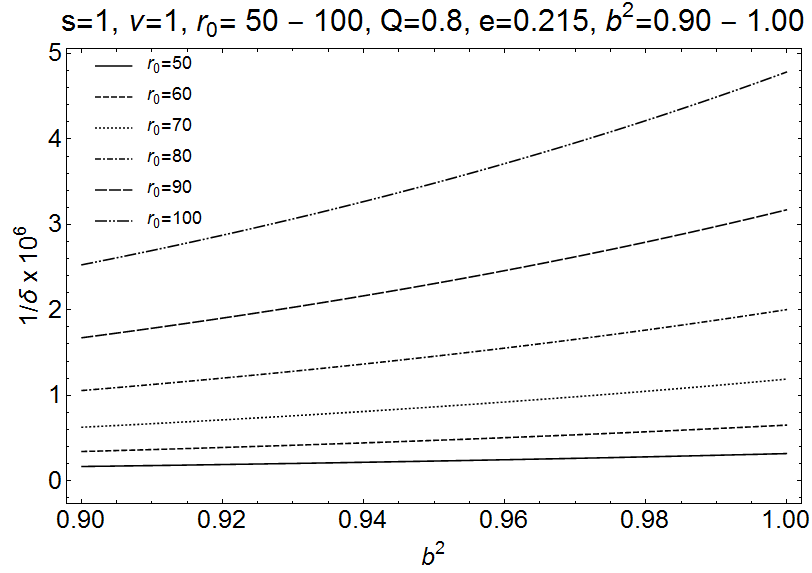} 
		\caption{Graph of $\tau$ as a function of $b^2$ for $r_0=50-100$ where we have fixed the other parameters as $s=1$, $\nu=1$, $Q=0.8$, $e=0.215$. i)As the mirror radius increases, the time scale increases. ii) As $b^2$ deacreases the time scale decreases. iii) When the mirror radius gets larger values the slope of the time scale increases for increasing $b^2$.} 
		\label{fig3:BHBradius:c} 
	\end{subfigure}
	\begin{subfigure}[b]{0.5\linewidth}
		\centering
		\includegraphics[width=0.96\linewidth]{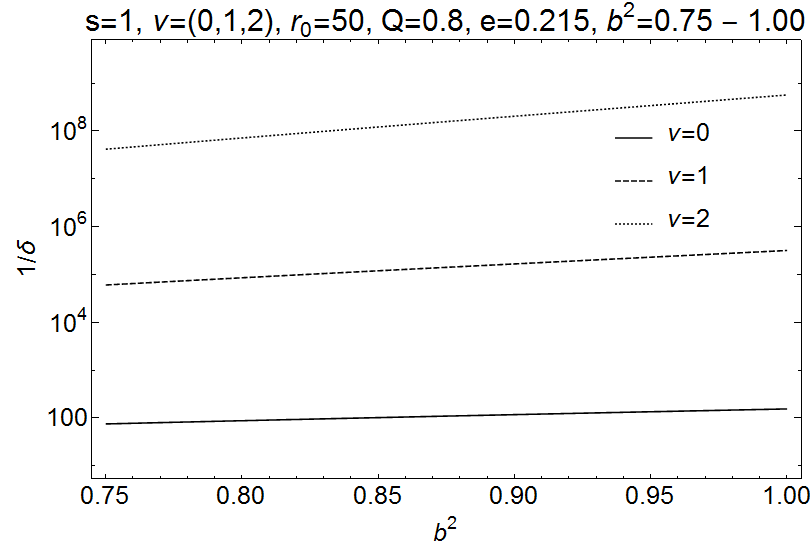} 
		\caption{Logarithmic graph of $\tau$ as a function of $b^2$ for the first three modes $\nu=0, 1, 2$ for black hole bomb. Similar to the figure (\ref{fig2:nodes:d}), the straight line corresponding to the fundamental mode, i.e. $\nu=0$, shows a little  change compared with the higher order modes as well. Note that the change in the values of time scales for each node is smaller compared to figure (\ref{fig2:nodes:d}).  } 
		\label{fig3:BHBnodes:d} 
	\end{subfigure} 
	\caption{Figure (\ref{fig3:BHBcond:a}) specifies the regime of the superradiant scattering and the variation of the threshold frequency with the gravitational monopole term $b^2=1-8\pi\eta^2$. As we have discussed in the text, the threshold frequency increases with increasing $\eta$. Figures (\ref{fig3:BHBcharge:b}), (\ref{fig3:BHBradius:c}), (\ref{fig3:BHBnodes:d}) corresponds to the different values of  black hole charge, mirror radius and modes respectively, shows the change of the time scale with respect to $b^2$. The values of the parameters in the graph are chosen such that the superradiance condition is satisfied.}
	\label{fig3:BHBdiff:parameters} 
\end{figure}

In order to better understand the effects of the monopole term, we again present several graphs  for this case as well. We observe that the behaviour of the time scale is very similar to the first case. The black hole charge $Q$ and frequency modes $\nu$ has similar graphs with different values, compared to the previous case. Figures (\ref{fig3:BHBradius:c}) and (\ref{fig2:cosmconst:c}) are also comparable since AdS space-time behaves as a reflecting box. The figure (\ref{fig3:BHBcond:a}) is different from its counterpart, namely,  figure (\ref{fig1:SupCond}), since the coupling of the gravitational monopole is different in each case. Therefore we see that  the threshold frequency is more sensitive in the changes of $b^2$ relative to the AdS case. Note that in figure (\ref{fig3:BHBcond:a}) we choose a small value of mirror radius $r_0$, which means we put the mirror closer to the black hole compared to the AdS radius $\ell$ in the first case.

Hence the results that we have obtained for both cases are quite similar. The main difference lies in the fact that monopole term affects the real part of the frequency modes of the black hole bomb by a factor $b^2$ but for RN-AdS case the factor is $b$ as the calculation procedure reveals, as a result, we may say that the chances of instability to occur is more likely in comparison with the AdS space. Another difference is the mode dependence of the instability time scale, since in the AdS case an explicit mode dependence exist with $b^{-2\nu}$ term, whereas there is no such dependence in the black hole bomb case as there is only $b^2$ term exists in this case. Hence the mode dependence of the black hole bomb  is only originated from the effect of the monopole on the horizons of the black hole. Hence, we see that, to obtain more accurate results concerning the comparison of superradiant instability in RN-AdS space-time with the black hole bomb in RN spacetime, a numerical analysis is also needed.

\section{Conclusion}

In this article, we have studied the dynamics of a massive, electrically charged scalar field in the background of a global monopole swallowed by an RN-AdS black hole space-time by investigating the charged and massive Klein-Gordon equation. Analyzing the asymptotic behavior of the scalar field near the horizon and far from it, we have discussed the effect of the monopole on the superradiance threshold frequency. We see that, since the monopole term increases the location of the outer horizon and this frequency depends on the electric potential of the horizon, the existence of the monopole decreases the electric potential and hence the threshold frequency. Therefore, a wave that leads to superradiant scattering for RN(-AdS) spacetime may not lead to a superradiant scattering in the presence of the monopole charge. Then we have exploited the asymptotic matching technique to inspect the stability conditions of both RN-AdS-monopole and RN-monopole black hole against charged scalar perturbations and found that global monopole affects the onset of instability in a negative way by coupling with the outer horizon of the black hole. Due to different couplings  to $b^2$ terms for both cases, the onset is affected more for the BH bomb case than the RN-AdS case. The time scale of the scalar field is also affected by global monopole and causes the instability to grow slower both in the RN-AdS-monopole and RN-monopole space-times due to the effect of the gravitational monopole on the outer horizon. We have presented several figures to better see the effect of the monopole in these black holes. As a result of this paper, we conclude that, the existence of a global monopole makes these black holes more stable against superradiance instability.

\section{Acknowledgements}
M. H. S. and O. D. are supported by Marmara University Scientific Research Projects Committee (Project No: FEN-C-YLP-150218-0055).

\end{document}